\documentstyle[12pt]{article}

\newcommand{\be}{\begin{equation}}
\newcommand{\ee}{\end{equation}}
\newcommand{\ba}{\begin{eqnarray}}
\newcommand{\ea}{\end{eqnarray}}
\newcommand{\baa}{\begin{eqnarray*}}
\newcommand{\eaa}{\end{eqnarray*}}
\newcommand{\bb}{\begin{thebibliography}}
\newcommand{\eb}{\end{thebibliography}}

\newcommand{\bi}[1]{\bibitem{#1}}
\newcommand{\lab}[1]{\label{#1}}
\newcommand{\re}[1]{(\ref{#1})}
\def\stackreb#1#2{\ \mathrel{\mathop{#1}\limits_{#2}}}

\newcounter{my}
\newcommand{\he}%
   {\stepcounter{equation}\setcounter{my}%
   {\value{equation}}\setcounter{equation}0%
   }%
\newcommand{\she}%
   {\setcounter{equation}{\value{my}}%
    }%

\begin{document}

%\hspace*{2mm}

\begin{center}
{\bf  THE FACTORIZATION METHOD, SELF-SIMILAR POTENTIALS \\[1mm]
  AND QUANTUM ALGEBRAS}%
\footnote{


Published in: {\it ``Special Functions 2000: Current Perspective and
Future Directions", Proc. NATO Advanced Study Institute
(Tempe, USA, May 29 -- June 9, 2000),
edited by J. Bustoz, M.E.H. Ismail and S.K. Suslov, Kluwer Academic
Publishers, Dordrecht, 2001, pp. 335--364.}
}
\\

\vspace{4mm}

V.P. SPIRIDONOV%

\medskip

{\it Bogoliubov Laboratory of Theoretical Physics, \\
JINR, Dubna,  Moscow region 141980, Russia } \\[5mm]
\end{center}

\begin{abstract}
The factorization method is a convenient operator language
formalism for consideration of certain spectral problems.
In the simplest differential operators realization it
corresponds to the Darboux transformations technique for
linear ODE of the second order. In this particular case
the method was developed by Schr\"odinger and became well
known to physicists due to the connections with quantum
mechanics and supersymmetry. In the theory of orthogonal
polynomials its origins go back to the Christoffel's theory
of kernel polynomials, etc. Special functions are defined
in this formalism as the functions associated with
similarity reductions of the factorization chains.

We consider in this lecture in detail the Schr\"odinger
equation case and review some recent developments in this
field. In particular, a class of self-similar potentials
is described whose discrete spectrum consists of a finite
number of geometric progressions. Such spectra are
generated by particular polynomial quantum algebras which
include $q$-analogues of the harmonic oscillator and
$su(1,1)$ algebras. Coherent states of these potentials
are described by differential-delay equations of
the pantograph type. Applications to infinite soliton
systems, Ising chains, random matrices, and lattice Coulomb gases
are briefly outlined.
\end{abstract}

%\bigskip\bigskip\

%\newpage
\tableofcontents
%\newpage
\section{Introduction}

   From the very beginning quantum mechanics served as a rich source of
good mathematical problems. It played a major role in the
development of the theory of generalized functions, functional analysis,
path integrals, to name a few. In the last two decades the ``quantum"
disease became so widespread in mathematics
that it is difficult to guarantee that a
randomly chosen mathematical term will not get its cousin
with such an adjective in the foreseeable future (if it does not have already).
The influence of quantum mechanics upon the theory of special
functions is also indispensable. A bright example is given by the
angular momentum theory which has lead to the Racah polynomials.
An aim of this lecture is to outline another fruitful interplay between
these two scientific fields inspired by the factorization method.

This method was suggested by Schr\"odinger as a convenient operator
language tool for working with quantum mechanical spectral problems
\cite{Sc1}. It was reformulated as a problem of searching of the
factorization chain solutions by Infeld \cite{Inf}.
The review \cite{IH} became a basic reference in this field.
A fairly recent revival of the interest to this method occurred due to
the discovery of its relation to the notion of supersymmetry (see, e.g.
the review \cite{GK} and references therein). The author himself turned to
this subject from the quantum chromodynamics due to an idea of a
generalization of the supersymmetric quantum mechanics \cite{RS}.
Although it was clear that special functions play an important role
in this formalism, it took some time to recognize that, heuristically,
special functions are defined in this approach as the functions appearing
from similarity reductions of the factorization chain (for a more precise
formulation, see the Appendix).

   From the mathematical side, a related techniques for solving linear
ordinary differential equations of the second order was proposed by
Darboux long ago \cite{Dar}. Its various generalizations are
referred to in the modern theory of completely integrable systems
as the Laplace, Darboux, B\"acklund, dressing, etc
transformations \cite{AS}; in the analysis of isomonodromic deformations
one deals with the Schlesinger transformations \cite{Ki}.
In the theory of special functions such
transformations appear as {\it contiguity relations}.
An important step in the development of the subject was performed
in a series of papers by Burchnall and Chaundy \cite{BC1},
which contain even some parts of the operator formulation of the
approach (``the transference"). For a completeness, let us mention also
the terms ``shift operator" and ``transmutation", which are used in
some other variations of the formalism. For a rigorous mathematical
treatment of some aspects of the Darboux transformations technique
or the factorization method, see \cite{Crum,D,F,Kr,Schm}.

As far as the discrete recurrence relations are concerned,
actually, it is the Christoffel's theory of kernel polynomials
that provides a first constructive approach of such kind to spectral
problems. This theory is based upon the simplest discrete analogue of
the Darboux transformations. A complementary part to this Christoffel's
transformation was found by Geronimus in \cite{Ger}. First applications of
the Schr\"odinger's factorization method to finite-difference equations
is given in \cite{Mi1}. Recent developments in this direction
are reviewed in \cite{SZ}. Let us remark that the same
techniques was rediscovered in the works on numerical
calculations of matrix eigenvalues. More precisely,
the well-known numerical $LR, QR, g$-algorithms, etc provide particular
instances of the chains of discrete Darboux transformations.
The literature on the taken subject is enormous, it is not possible to
describe all its branches. The list of references given at the
end of this manuscript is not complete, it contains mainly the papers
encountered by the author during his own work (additional
lists can be found in \cite{S2,SZ}).

Despite of a vast variety of existing constructions, part of
which was just mentioned, we limit ourselves in this lecture
to the simplest possible case based upon the stationary
one-dimensional Schr\"odinger equation
\begin{equation}\label{schr-eq}
L\psi(x)=-\psi_{xx}(x)+u(x)\psi(x)=\lambda\psi(x),
\end{equation}
describing the motion of a non-relativistic particle on the line
$x\in \bf R$ in the potential field $u(x)$, which is assumed to be
bounded from below. The operator $L$ is called
the Hamiltonian or the Schr\"odinger operator. For convenience, the
particle's mass variable and the Planck's constant $\hbar$ are removed
by rescalings of the coordinate $x$ and the energy $\lambda$.
The equation (\ref{schr-eq}) describes an eigenvalue problem for $L$
and the eigenvalues (or the permitted bound states energies of the quantum particle)
are determined from the condition that the modulus of the wave function
$\psi(x)$ is square integrable, $\psi(x)\in L^2({\bf R})$.

Depending on the physical situation, the real line $\bf R$ may be replaced
by an interval with an appropriate boundary conditions upon $\psi(x)$.
However, if one is interested not in the spectra themselves, but
in the differential operators $L$ with some formal properties, then
it is convenient to take $x, \lambda\in \bf C$.

\section{The factorization method}

Let us factorize the second order differential operator $L=-d^2/dx^2+u(x)$
as a product of two first order ones up to some real constant $\lambda_0$:
\begin{equation}\label{fact}
L=A^+A^-+\lambda_0, \qquad A^\pm=\mp d/dx +f(x).
\end{equation}
The function $f(x)$ is a solution of the Riccati equation
$f^2(x)-f_x(x)+\lambda_0=u(x)$. Substitution of the ansatz
$f(x)=-\phi_{0,x}(x)/\phi_0(x)$ shows that $-\phi_{0,xx}(x)+u(x)\phi_0(x)
=\lambda_0\phi_0(x)$, i.e. $\phi_0(x)$ is a solution of the original
Schr\"odinger equation (\ref{schr-eq}) for $\lambda=\lambda_0$.

If $f(x)$ is a smooth function, then $A^+$ is a hermitean conjugate
of $A^-$ in $L^2({\bf R})$ and $L$ is a self-adjoint operator.
Under these circumstances $\lambda_0$
cannot be bigger than the smallest eigenvalue of $L$. Suppose that
$\lambda_0$ is the smallest eigenvalue of $L$, and let $\psi_0(x)$ be the
corresponding eigenfunction. It is well known that $\psi_0(x)$ is
nodeless and may be normalized to have the unit norm, $||\psi_0||^2=
\int_{-\infty}^\infty|\psi_0(x)|^2dx=1.$
Then,
$$
\phi_0(x)=a\psi_0(x)+b\psi_0(x)\int^x\frac{dy}{\psi_0^2(y)},
$$
where $a,b$ are arbitrary constants,
is the general solution of the equation $L\phi_0=\lambda_0\phi_0$.
    For $b\neq 0$, the resulting $f(x)$ is singular at some point
and the operators $A^\pm$ in (\ref{fact}) are not well defined.
Let us exclude this situation, i.e. set $a=1, b=0$.

Using the lowest eigenvalue eigenfunction of $L$, one can always
factorize $L$ as described in (\ref{fact}) with the well-defined
operators $A^\pm$. Vice versa, if one manages
to find the factorization (\ref{fact}) such that the zero mode $\psi_0$
of the operator $A^-$, $A^-\psi_0=0$, belongs to $L^2({\bf R})$
and $f(x)$ is not singular, then $\lambda_0$ is the lowest eigenvalue and
\begin{equation}\label{zmode}
\psi_0(x)=\frac{e^{-\int_{0}^xf(y)dy}}
{\left(\int_{-\infty}^\infty e^{-2\int_{0}^x f(y)dy}dx\right)^{1/2}}
\end{equation}
is the corresponding normalized eigenfunction.

Let us define now a new Schr\"odinger operator $\tilde L$ by
the permutation of the operator factors in (\ref{fact})
\begin{equation}\label{new-op}
\tilde L=A^-A^+ + \lambda_0,
\end{equation}
whose potential has the form $\tilde u(x)=f^2(x)+f_x(x)+\lambda_0$.
Evidently, one has the intertwining relations
\begin{equation}\label{inter}
LA^+=A^+\tilde L, \qquad A^-L=\tilde LA^-,
\end{equation}
playing the key role in the formalism.
  From (\ref{inter}) one deduces that if $\psi(x)$ satisfies (\ref{schr-eq}),
then the functions $\tilde\psi=A^-\psi$ provide formal eigenfunctions
of $\tilde L$. Indeed,
\begin{equation}\label{interpsi}
\tilde L\tilde\psi=\tilde L (A^-\psi)=A^-(L\psi)=\lambda(A^-\psi).
\end{equation}
Actually, this gives general solutions of the
differential equation $\tilde L\tilde\psi=\lambda\tilde\psi$
for all $\lambda$, except of the point $\lambda=\lambda_0$,
where the zero mode of $A^-$ is located. This problem is curable
since one can find the general solution of the equation
$\tilde L\tilde\phi_0=\lambda_0\tilde\phi_0$ separately:
\begin{equation}\label{gensol2}
\tilde\phi_0(x)= \frac{g}{\psi_0(x)}+\frac{e}{\psi_0(x)}\int^x\psi_0^2(y)dy,
\end{equation}
where $g,e$ are arbitrary constants and $\psi_0$ is the lowest eigenvalue
eigenfunction of $L$ (note that $A^+\psi_0^{-1}=0$).

Denote as $\lambda_n, \tilde \lambda_n$ and $\psi_n,$ $\tilde\psi_n=A^-\psi_n$
discrete eigenvalues and corresponding eigenfunctions of $L$ and $\tilde L$
respectively. From (\ref{interpsi}) it follows that the spectra
$\lambda_n$ and $\tilde \lambda_n$ almost coincide
$\tilde\lambda_n=\lambda_n, n=1,2,\dots$. The only difference that may
occur in these spectra concerns the zero mode of the operator $A^-$.
In fact, the point $\lambda=\lambda_0$ does not belong to the
spectrum of $\tilde L$. Indeed, since $\psi_0\in L^2({\bf R})$,
it follows that $\psi_0^{-1}\notin L^2({\bf R})$,
and, as a result, for any $g,e$ the function
(\ref{gensol2}) cannot be normalizable.

Thus, $\lambda_1$ is the lowest eigenvalue of $\tilde L$
and the point $\lambda_0$ was ``deleted" from the spectrum of $L$.
Repeating the same procedure once more, i.e. taking
$\tilde L=\tilde A^+\tilde A^-+\lambda_1$ and permuting the
operator factors, one can delete the point $\lambda=\lambda_1$, etc.
This procedure allows one to remove an arbitrary number of smallest
eigenvalues of $L$. Often it is much easier to find the smallest
eigenvalue of a given operator than the other ones.
If the lowest eigenvalue of $\tilde L$ is determined separately
by some means, then, by construction, it will coincide
with the second eigenvalue of $L$, etc. This observation is the central
one in the factorization method \cite{Sc1} since it reduces
the problem of finding complete discrete spectrum of a taken operator to
the problem of finding lowest eigenvalues of a sequence of operators
built from $L$ by the ``factorize and permute" algorithm.

One can invert the procedure of a deletion of the smallest eigenvalue.
Namely, a given $\tilde L$ with known lowest eigenvalue
$\lambda_1$ may be factorized as $\tilde L=A^-A^++\lambda_0$,
with $\lambda_0<\lambda_1$. If the zero mode of $A^-$ is normalizable,
then the operator $L=A^+A^-+\lambda_0$ has the same spectrum as $\tilde L$ with an
additional {\it inserted} eigenvalue at an {\it arbitrary} point
$\lambda= \lambda_0$. If one factorizes $L$ (or $\tilde L$) in such
a way that $f(x)$ is a non-singular function, but none of the zero modes of
the operators $A^\pm$ are normalizable, then the discrete spectra of $L$
and $\tilde L$ coincide completely (an isospectral situation).

There are more complicated possibilities for changing spectral data of a
given Schr\"odinger operator $L$. E.g., if the first factorization
is ``bad", in the sense that $\tilde L$ has a singular potential
and considerations given above are not valid, then
one may demand that after a number of additional refactorizations one gets
a well defined self-adjoint Schr\"odinger operator. In this way
one can delete not only the lowest eigenvalues of $L$ or to insert the
new ones, but delete or insert a bunch of spectral points above the smallest
one \cite{Kr}. In particular, two step refactorization procedure allows one
to imbed an eigenvalue into the continuous spectrum, etc \cite{D}.

Let us give now a ``discrete time" formulation of the construction.
Denote $L\equiv L_j$ and $\tilde L\equiv L_{j+1}$ and take $j\in {\bf Z}$
($j$ may be treated as a continuous parameter and we could
denote $\tilde L \equiv L_{j-1}$ --- all this is a matter of agreement).
This gives an infinite sequence of Schr\"odinger operators
$L_j=-d^2/dx^2+u_j(x)$ with formal factorizations
\begin{equation}
L_j=A_j^+A_j^- + \lambda_j,\qquad A_j^\pm=\mp\;d/dx + f_j(x).
\label{fact2}\end{equation}
Neighboring $L_j$ are connected to each other via the
{\t abstract factorization chain}
\begin{equation}\label{fchain}
L_{j+1}=A_{j+1}^+A_{j+1}^-+\lambda_{j+1}=A_j^-A_j^++\lambda_j.
\end{equation}
Intertwining relations take the form
$$
A_j^-L_j=L_{j+1}A_j^-, \qquad L_jA^+_j=A_j^+L_{j+1}.
$$
Substituting explicit forms of $A^\pm_j$ into (\ref{fchain})
one gets a differential-difference equation upon $f_j(x)$:
\begin{equation}\label{chain}
(f_j(x)+f_{j+1}(x))_x+f_j^2(x)-f_{j+1}^2(x)=
\mu_j\equiv \lambda_{j+1}-\lambda_j.
\end{equation}
This chain was derived in \cite{Inf} and the problem of
searching ``exactly solvable" spectral problems was formulated as
a problem of the search of solutions of (\ref{chain}) such that
the points $\lambda_j$ define the discrete spectrum of an operator $L$,
say, $L\equiv L_0$. E.g., one may try to find solutions of
the equation (\ref{chain}) $f_j(x)$ in the form of power series in $j$
($\lambda_j$ are considered as unknown functions of $j$).
As shown in \cite{Inf,IH} the finite term expansion occurs iff
$f_j(x)=a(x)j+b(x)+c(x)/j$, where $a, b, c$ are some elementary
functions of $x$. This leads to the $_2F_1$ hypergeometric function
and well known ``old" exactly solvable potentials of quantum mechanics.

As evident from the construction,
the constants $\lambda_j, j\geq 0,$ determine the smallest eigenvalues of
$L_j$ under the condition that the zero modes of $A_j^-$ are normalizable
and $f_j(x)$ are not singular.
In general, the $j\to j+1$ transitions may describe all three possibilities
--- removal or insertion of an eigenvalue and isospectral transformations
(sometimes it is convenient to
parameterize inserted eigenvalues as $\lambda_0>\lambda_1 >\dots>\lambda_n$).

  For a positive integer $n$, let us introduce the operators
$$
M_j^-=A_{j+n-1}^-\cdots A_{j+1}^-A_j^-, \qquad
M_j^+=A_j^+A_{j+1}^+\cdots A_{j+n-1}^+.
$$
The intertwining relations
$$
L_{j+n}M_j^-=M_j^-L_j, \qquad M_j^+L_{j+n}=L_jM^+_j
$$
guarantee that for almost all $\lambda$ solutions of the equations
$L_j\psi^{(j)}=\lambda\psi^{(j)}$
and $L_{j+n}\psi^{(i+n)}=\lambda\psi^{(j+n)}$ are related to each other
as $\psi^{(j+n)}\propto M_j^-\psi^{(j)}$ and
$\psi^{(j)}\propto M^+_j\psi^{(j+n)}$. As a result, the product
$M^+_jM^-_j$ should commute with $L_j$ and $M^-_jM^+_j$ should
commute with $L_{j+n}$. A simple computation leads to the equalities
\be
M_j^+M_j^-=\prod_{k=0}^{n-1}(L_j-\lambda_{j+k}), \qquad
M_j^-M_j^+=\prod_{k=0}^{n-1}(L_{j+n}-\lambda_{j+k}).
\lab{polrel}\ee
Let $\psi^{(j)}(x)\in L^2({\bf R})$ be an eigenfunction of $L_j$
with the eigenvalue $\lambda$ and some finite norm $||\psi^{(j)}||$.
If we set $\psi^{(j+n)}=M_j^-\psi^{(j)}$, then
$$
||\psi^{(j+n)}||^2=(\lambda-\lambda_j)\cdots(\lambda-\lambda_{j+n-1})\,
||\psi^{(j)}||^2.
$$
Zeroes on the r.h.s. for $\lambda=\lambda_k$ for some $k$ indicate that the
corresponding eigenvalues were deleted from the spectrum.
Under the condition that zero modes of $A^-_j, j=0, \dots, n-1,$ are
normalizable and nodeless, $A^-_j\psi_0^{(j)}=0$, $||\psi_0^{(j)}||=1$,
one finds that the functions
$$
\psi^{(0)}_n(x)=\frac{A^+_0\cdots A^+_{n-1} \psi_0^{(n)}(x)}
{\sqrt{(\lambda_n-\lambda_{n-1})\cdots(\lambda_n-\lambda_0)}}
$$
define the unit norm eigenfunctions of $L_0$ with the eigenvalues
$\lambda_n$:
$L_0\psi^{(0)}_n=\lambda_n\psi^{(0)}_n$ and $||\psi_n^{(0)}||=1$.

The main advantage of the factorization method consists in its pure operator
language formulation. One may replace the Schr\"odinger operator by any
other (higher order differential, finite-difference, integral, etc)
operator $L$ admitting factorizations into the well defined operator factors
of a simpler nature $A^\pm$ (for some applications they are not
necessarily hermitean conjugates of each other).
In all cases one deals with the
abstract operator factorization chain (\ref{fchain})
with an appropriate interpretation of the constants  $\lambda_j$.
It should be noted that this method does not have straightforward
generalizations to the multidimensional spectral problems, only
some of its features are preservable, see e.g. \cite{ABI}.

\section{Supersymmetry}

Supersymmetry is a symmetry between bosonic and fermionic degrees
of freedom of particular physical systems. The corresponding
symmetry algebras are distinguished from the standard Lie algebras
by the presence of anticommutator relations. The simplest superalgebra
is realized upon two neighboring operators in the factorization
chain. Define the matrix operators (supercharges)
$$
Q^+=\left(\begin{array}{cc}
 0 & A^+ \\
 0   & 0
\end{array}\right),\quad
Q^-=\left(\begin{array}{cc}
 0 & 0 \\
 A^-   & 0
\end{array}\right),
$$
where $A^\pm$ are the factorization operators defined in (\ref{fact}).
They form the following algebra (see, e.g. \cite{GK})
\begin{equation}\label{susy}
\{Q^+,Q^-\}=H, \qquad [H,Q^\pm]=(Q^\pm)^2=0,
\end{equation}
where the Hamiltonian
$$
H=\left(\begin{array}{cc}
 L - \lambda_0 & 0 \\
 0   & \tilde L-\lambda_0
\end{array}\right)=-\frac{d^2}{dx^2}+f^2(x)-f_x(x)\sigma_3
$$
describes a particle with spin 1/2 (a fermionic variable) on
the line in an external magnetic field, $\sigma_3$ is the
Pauli matrix. In terms of the
hermitean supercharges $Q_1=Q^++Q^-, Q_2=(Q^+-Q^-)/i$, the
algebra takes the form $\{Q_i,Q_j\}=2H\delta_{ij},\; [H,Q_i]=0$.

As a formal consequence of the algebra (\ref{susy}) the
discrete spectrum of $H$
is doubly degenerate (such degeneracies are characteristic
to the supersymmetry) with possible exception of the lowest
eigenvalue which cannot be negative (this is another constraint upon
the supersymmetric systems).

This simple construction was generalized in \cite{RS} to a
symmetry between particles with parastatistics.
The corresponding symmetry algebras are polynomial
in the generators. E.g., in the simplest case involving a
parafermion of the second order (or the spin 1 particle)
$H$ is given by a $3\times 3$ diagonal matrix with the
entries $L_0, L_1, L_2$, the corresponding symmetry generators being
$(Q^+)_{ij}=A_i^+\delta_{i,j-1}$ and its conjugate.
  For the hermitean charges $Q_{1,2}$ defined as above, one has
now cubic relations
\begin{equation}\label{psusy}
Q_i(\{Q_j,Q_k\}-2H\delta_{jk})+\mbox{cyclic perm. of }i,j,k  =0,
\qquad [H,Q_i]=0.
\end{equation}
The discrete spectrum of $H$ is now triply degenerate with possible
exception of two smallest eigenvalues.

One can go further and propose other modifications of the algebra
(\ref{susy}). For instance, one can $q$-deform it \cite{S0}:
$$
Q^+Q^-+q^2Q^-Q^+=H,\qquad (Q^\pm)^2=0,\qquad HQ^\pm=q^{\pm 2}Q^\pm H.
$$
This lifts the degeneracy of spectra, but creates a nontrivial
scaling relation between spectral points. As a result, it opens
a way for building various $q$-harmonic oscillator models.
Another construction proposed in \cite{AIS}
refers to a polynomials generalization of (\ref{susy}). It appears
after a reduction of the algebra (\ref{psusy}) and similar higher order
polynomial relations to a two-dimensional subspace of eigenfunctions:
$$
\{Q^+, Q^-\}= P_n(H), \qquad [H, Q^\pm]=(Q^\pm)^2=0,
$$
where $P_n(H)=\prod_{k=0}^{n-1}(H-\lambda_k)$ and
$$
Q^+=\left(\begin{array}{cc}
 0 & A_0^+\cdots A_{n-1}^+ \\
 0   & 0
\end{array}\right), \qquad
Q^-=\left(\begin{array}{cc}
 0 & 0 \\
A_{n-1}^-\cdots A_0^-  & 0
\end{array}\right), \qquad
H=\left(\begin{array}{cc}
 L_0 & 0 \\
 0   & L_n
\end{array}\right).
$$
This is a pure supersymmetry again in the sense of a symmetry
between bosons and fermions. However, the consequences
are quite different from the standard case (\ref{susy}).
In particular, there are no such severe constraints
upon the smallest eigenvalue of $H$ (the vacuum energy).

All these general constructions are useful because they contain
new concepts providing nonstandard viewpoints upon physical
systems and new mathematical tools for their exploration.
Special functions emerge when one starts to work with
particular ``exactly solvable" models with such symmetries.

\section{Darboux transformations}

Let us describe the Darboux transformations technique in its
modern appearance. Consider compatibility conditions of
the following three linear equations
\begin{equation}\label{eq1}
L_j\psi^{(j)}(x,t) = \lambda\psi^{(j)}(x,t), \qquad
L_j \equiv -\partial_x^2 + u_j(x,t),
\end{equation}
\begin{equation}\label{eq2}
\psi^{(j+1)}(x,t) = A_j^-\psi^{(j)}(x,t), \qquad
A_j^-\equiv \partial_x + f_j(x,t),
\end{equation}
\begin{equation}\label{eq3}
\psi_t^{(j)}(x,t)=B_j\psi^{(j)}(x,t), \qquad
B_j\equiv -4\partial_x^3+6u_j(x,t)\partial_x+3u_{j,x}(x,t),
\end{equation}
where $t$ is some additional continuous parameter (``evolution time") and
$u_j(x,t), f_j(x,t)$ are some free functions.
The compatibility condition of (\ref{eq1}) and (\ref{eq2}) generates the
intertwining relation $A_j^-L_j = L_{j+1}A_j^-$, the resolution of which
yields the constraints $u_j = f_j^2 - f_{j,x} + \lambda_j,$
$u_{j+1} = u_j + 2f_{j,x}$,
where $\lambda_j$ is an integration constant. Thus one arrives again
to the infinite chain of nonlinear differential-difference equations
(\ref{chain}) and the formal factorizations (\ref{fact2}).

The compatibility condition of the equations (\ref{eq1}) and (\ref{eq3})
yields the celebrated Korteweg-de Vries (KdV) equation:
\begin{equation}\label{kdv}
u_{j,t}(x,t)-6u_j(x,t) u_{j,x}(x,t)+u_{j,xxx}(x,t)=0,
\end{equation}
a completely integrable Hamiltonian system with an infinite
number of degrees of freedom \cite{AS}. If we express $u_j(x,t)$
through $f_j(x,t)$ in the equation (\ref{kdv}), then it takes the form
$$
(-\partial_x+2f_j)\, V_j(x,t)=0, \qquad V_j(x,t)\equiv
f_{j,t}-6(f_j^2+\lambda_j)f_{j,x}+f_{j,xxx}.
$$
It appears that the compatibility conditions of this equation
with the Darboux transformation (\ref{eq2}) imposes an additional
non-trivial constraint $V_j=0$, which is called the modified KdV
equation.

A substitution of $f_j = -\phi_{0,x}^{(j)}/\phi_0^{(j)}$ into
the relation $u_j=f_j^2-f_{j,x}+\lambda_j$ yields
$-\phi_{0,xx}^{(j)}+u_j\phi_0^{(j)}=\lambda_j\phi_0^{(j)}$
and the Darboux transformation (\ref{eq2}) takes the form:
$$
\psi^{(j+1)} = (\phi_0^{(j)}\psi^{(j)}_x - \phi_{0,x}^{(j)}\psi^{(j)})/
\phi_0^{(j)},
$$
where $\phi_0^{(j)}$ is a particular solution of the $j$-th equation in
the sequence (\ref{eq1}) for $\lambda=\lambda_j$.

Let $L_j\phi_k^{(j)}=\lambda_{j+k}\phi_k^{(j)}$, i.e. let
$\phi_k^{(j)}(x,t)$ are formal eigenfunctions of $L_j$ with the
eigenvalues $\lambda_{j+k}$
equal to constants of integration mentioned above for all $k$.
Denote as $W(\phi_1,\dots,\phi_n)=\det(\partial_x^{i-1}\phi_k)$
the Wronskian of a set of functions $\phi_k$. Using the identities
$W(\xi(x)\phi_1,\xi(x)\phi_2)=\xi^2(x)W(\phi_1,\phi_2)$ and
$$
W\left(W(\phi_1,\dots,\phi_n,\xi_1), W(\phi_1,\dots,\phi_n,\xi_2)\right)
=W(\phi_1,\dots,\phi_n)W(\phi_1,\dots,\phi_n,\xi_1,\xi_2),
$$
one can show that \cite{Crum}
\begin{equation}\label{crum}
u_{j+n}(x,t)=u_j(x,t)-2\partial_x^2\log
W(\phi_0^{(j)},\dots,\phi_{n-1}^{(j)}),
\end{equation}
\begin{equation}\label{crum2}
f_{j+n}(x,t)=-\partial_x\log \frac{W\left(\phi_0^{(j)},\dots,
\phi_n^{(j)}\right)}{W\left(\phi_0^{(j)},\dots,\phi_{n-1}^{(j)}\right)},
\end{equation}
\begin{equation}\label{crum3}
\psi^{(j+n+1)}(x,t)=\frac{W\left(\phi_0^{(j)},\dots,\phi_n^{(j)},\psi^{(j)}\right)}
{W\left(\phi_0^{(j)},\dots,\phi_n^{(j)}\right)}=
(\partial_x+f_{j+n})\cdots(\partial_x+f_j)\psi^{(j)}(x,t).
\end{equation}
These formulae give an explicit representation of the transformed potentials
$u_{j+n}$ in terms of the initial Schr\"odinger equation solutions
$\phi^{(j)}_k$.

Since the parameter $t$ is a dummy variable during the Darboux
transformations, in fact one deals simultaneously with different solutions
of the KdV equation: if $u_j(x,t)$ satisfies (\ref{kdv}), the same
is true for $u_{j+n}(x,t)$. This gives a way for building
new complicated explicit solutions of the KdV equation starting from
the simple ones. For instance, one may take $u_0(x,t)=0$ which gives
$\phi_m^{(0)}(x,t)=a_me^{\kappa_mx-4\kappa_m^3t}+b_me^{-\kappa_mx+4\kappa_m^3t}$,
and for some special choice of the signs of $b_m/a_m$ the potential
$u_n(x,t)$ becomes a nonsingular reflectionless potential with
$n$ discrete spectrum points $\lambda_m= -\kappa_m^2<0$. It
defines the famous $n$-soliton solution of the KdV equation.

In the formulae given above one may permute any pair of the $\phi_k^{(j)}$
functions and the final result is not changed, i.e. such a permutation
of the Darboux transformations is a symmetry of the factorization chain
(\ref{chain}). In the simplest case, it has the following explicit
form \cite{Ad}:
$$
\tilde f_k=f_k-\frac{\lambda_{k+1}-\lambda_k}{f_{k+1}+f_k},\quad
\tilde f_{k+1}=f_{k+1}+\frac{\lambda_{k+1}-\lambda_k}{f_{k+1}+f_k},\quad
\tilde \lambda_k=\lambda_{k+1},\quad
\tilde \lambda_{k+1}=\lambda_k,
$$
all other $f_j(x), \lambda_j$ staying intact for $j\neq k,k+1$.
This discrete symmetry may be ``discretized" further by passing to
the discrete Schr\"odinger equation (or the three term recurrence relation
for orthogonal polynomials) and the corresponding discrete Darboux
transformations \cite{S4}.

The well known KdV tau-function is introduced as $u_j(x,t)=-2\partial_x^2
\log\tau_j(x,t)$. In its terms relations (\ref{crum}), (\ref{crum2})
are rewritten as $\tau_{j+n}=W(\phi_0^{(j)},\dots,\phi_{n-1}^{(j)})\tau_j$,
$f_j=-\partial_x\log \tau_{j+1}/\tau_j$. Introducing the variables
$\rho_j=-\partial_x\log\tau_j$, one can write $u_j=2\rho_{j,x}$ and
$f_j=\rho_{j+1}-\rho_j$. As a result, the relation between
$u_j$ and $f_j$ yields the equation
$$
(\rho_{j+1}+\rho_j)_x-(\rho_{j+1}-\rho_j)^2=\lambda_j,
$$
which starts to play the role of the factorization chain. The function
$\tau_j$ is a very convenient object since its zeros in $x$ correspond to
poles of the potential.

\section{Operator self-similarity and quantum algebras}

Let us turn now to the problem of searching particular solutions
of the operator factorization chain (\ref{fchain}). At first glance
it is not clear how to proceed, but the harmonic oscillator problem ---
a base model for the whole quantum mechanics and quantum field theory ---
provides a guiding idea. One has to try to form from the factorization
operators $A^\pm_j$ some nontrivial symmetry algebras. In fact, the
relations (\ref{polrel}) look already as defining relations of some algebra,
but they are not closed --- the relations between the eigenvalue
problems  for $L_j$ and $L_{j+n}$ are too weak (they are valid for
any starting potential $u_j(x)$). In order to close the system, one has to
assume that there is an additional relation between $L_j$ and $L_{j+n}$,
say $L_{j+n}=g(L_j)$,
which would force the operators $M_j^\pm$ to map eigenfunctions of
a taken operator $L_j$ onto themselves.

In the simplest case one demands that the sequence of
Hamiltonians $L_j$ is periodic: $L_{j+N}=L_j$ for some period $N>0$.
As a result, $M^\pm_j$ commute with $L_j$, $[M^\pm_j,L_j]=0$. This is a
remarkable fact since the existence of additional conserved
quantities may simplify solution of the eigenvalue
problem for $L_j$. In the differential operator realization of $A^\pm_j$
this leads to the commuting differential operators \cite{BC1}.
There is a generalization of this pure periodicity condition to the
periodicity up to a twist condition
$L_{j+N}=UL_jU^{-1}$, where $U$ is some invertible operator.
This leads again to commuting operators $[B^\pm_j,L_j]=0$, where
$B^+_j=M^+_jU,\; B^-_j= U^{-1}M_j^-$,
but now there is an essential additional freedom in the choice of $U$.

Another possible ``closure" or a reduction of the sequence of operators
$L_j$ consists in the requirement of their periodicity up to a constant
shift and a twist, $L_{j+N}=UL_jU^{-1}+\mu$, where $\mu$ is a
constant. This results in the ladder relation
$[L_j,B^\pm_j]=\pm \mu B^\pm_j$ and the operator identities
$$
B^+_jB^-_j=\prod_{k=0}^{N-1}(L_j-\lambda_{j+k}),\qquad
B^-_jB^+_j=\prod_{k=0}^{N-1}(L_j+\mu-\lambda_{j+k}),
$$
where $B_j^\pm$ operators are defined as in the previous case.
Denoting $B^\pm\equiv B_0^\pm, L\equiv L_0$, one can form a
polynomial algebra
\begin{equation}\label{pol}
[L,B^\pm]=\pm\mu B^\pm,\qquad [B^+,B^-]=P_{N-1}(L),
\end{equation}
where $P_{N-1}(x)$ is a polynomial of the degree $N-1$ in $x$.
  For a representation theory of such algebras, see e.g. \cite{Sm}.
Note that for $N=1$ this is the Heisenberg-Weyl or the harmonic oscillator
algebra, and for $N=2$ it coincides with the $su(1,1)$ algebra.

Quantum algebras, or $q$-analogues of the algebras (\ref{pol})
appear from the following operator self-similarity constraint
imposed upon the chain (\ref{fchain}):
\begin{equation}\label{q-closure}
L_{j+N}=q^2UL_jU^{-1}+\mu.
\end{equation}
When the numerical factor $q^2\neq 1$, one can remove $\mu$ by the uniform
shift $L_j\to L_j+\mu/(1-q^2)$. Therefore we
assume below that $\mu=0$. Substitute (\ref{q-closure})
with $\mu=0$ into (\ref{polrel}). Then the operators $L=L_0$,
$B^+=M^+_0U,\; B^-= U^{-1}M_0^-$ satisfy the following identities
\begin{equation}\label{q-alg}
LB^\pm=q^{\pm 2}B^\pm L,\qquad
B^+B^-=\prod_{j=0}^{N-1}(L-\lambda_j),\qquad
B^-B^+=\prod_{j=0}^{N-1}(q^2L-\lambda_j).
\end{equation}
  For $N=1$ these relations provide a realization of the $q$-harmonic
oscillator algebra
\begin{equation}\label{qosc}
B^-B^+-q^2B^+B^-=\rho, \qquad [B^\pm,\rho]=0,
\end{equation}
with $\rho=\lambda_0(q^2-1)$. One can set $\rho=1$ by taking the
normalization condition $\lambda_0=1/(q^2-1)$.
Such a $q$-analogue of the Heisenberg-Weyl algebra was encountered in physics
long ago \cite{Coo,FB}. In the modern times it became quite popular due
to the inspirations coming from the quantum groups, see e.g. \cite{Mac}.
  For $N=2$ relations (\ref{q-alg}) determine a particular $q$-analogue of
the conformal algebra $su(1,1)$ admitting the Hopf algebra
structure, etc \cite{S0,S2}.

\section{Self-similar potentials}

Consider the Schr\"odinger equation realizations of the
algebraic relations described in the previous section.
Let us start from the closure $\tilde L=ULU^{-1}+\mu$,
where $L=-\partial_x^2+u(x)$ is the Schr\"odinger operator
in the notations of Sect. 2 and $U$ is a translation operator
$Uf(x)=f(x+a)$. As a result,
the operators $B^\pm$ take the form $B^+=(-\partial_x+f(x))U$,
$B^-=U^{-1}(\partial_x+f(x))$ and one gets the Heisenberg-Weyl
algebra $[L,B^\pm]=\pm\mu B^\pm, [B^-,B^+]=\mu$. The potential
entering $L$ is defined as $u(x)=f^2(x)-f_x(x)+\lambda_0$,
where $f(x)$ satisfies the following
nonlinear differential-delay equation
\begin{equation}\label{nonlocal}
\left(f(x)+f(x+a)\right)_x+f^2(x)-f^2(x+a)=\mu.
\end{equation}
  For $a=0$ one gets from this equation the standard harmonic
oscillator model $f(x)\propto x, u(x)\propto x^2$, which is related to
the Hermite polynomials. Although for $a\neq 0$ and $\mu=0$ the author
has found a meromorphic solution of (\ref{nonlocal}) in terms of
the Weierstrass $\cal P$-function
$$
f(x)=-\frac{1}{2}\,\frac{{\cal P}'(x-x_0)-{\cal P}'(a)}
{{\cal P}(x-x_0)-{\cal P}(a)},
$$
in general it is quite difficult to build its solutions
analytic in some region, especially for
$\mu\neq 0$. Let us remark, that the equation (\ref{nonlocal}) may
be derived also after imposing the constraint $f_{j+1}(x)=f_j(x+a)$
upon the chain (\ref{chain}).

Turn now directly to the relations (\ref{q-alg}). In this case
one can take $U$ as the dilation (or $q$-difference) operator:
$Uf(x)=|q|^{1/2}f(qx)$. For real $q\neq 0$ this is a unitary operator
$U^\dagger=U^{-1}$. Taking $L\equiv L_0,$
$B^-\equiv U^{-1}(\partial_x+f_{N-1})\cdots(\partial_x+f_0),$
$B^+\equiv (-\partial_x+f_0)\cdots(-\partial_x+f_{N-1})U$
one realizes the identities (\ref{q-alg}), provided $f_j(x)$ satisfy the
following system of nonlinear differential-delay equations:
$$
(f_0(x)+f_1(x))_x+f_0^2(x)-f_1^2(x)=\mu_0, \quad \dots\dots
$$
\begin{equation}\label{q-pereq}
(f_{N-1}(x)+qf_0(qx))_x+f_{N-1}^2(x)-q^2f_0^2(qx)=\mu_{N-1}.
\end{equation}
Equivalently, these equations appear from the chain (\ref{chain})
after imposing the following constraints
\begin{equation}
f_{j+N}(x)=qf_j(qx),\qquad \mu_{j+N}=q^2\mu_j,
\label{q-per}\end{equation}
having a simple group-theoretical interpretation.

One easily arrives at (\ref{q-per}) using the Lie's symmetry reduction
technique (for a differential context of this theory, see e.g. \cite{Mi2},
and for an extension to difference equations, see e.g. \cite{LW,Mae}).
   First, one notices that if $f_j(x), \;\mu_j$ are solutions of the
chain (\ref{chain}), then their discrete scaling transformation
$qf_j(qx), \; q^2\mu_j$ gives solutions of (\ref{chain}) as well.
Analogously, a shift in the numeration
$f_j(x)\to f_{j+N}(x),\; \mu_j\to\mu_{j+N}$ maps solutions of (\ref{chain})
into solutions. Let us consider a set of {\it self-similar}
solutions of (\ref{chain}) which is invariant under a combination
of these two symmetries. E.g., demanding that these two transformations
are equivalent to one another, we arrive to (\ref{q-per}).
  For $N=1$, this reduction may be rewritten as
$f_j(x)=q^jf_0(q^jx),\; \lambda_j=q^{2j}\lambda_0$ in which form it
was first met in \cite{Sh}. The general $q$-periodic reduction (\ref{q-per})
was found by the author \cite{S0} in an attempt to $q$-deform the
parastatistical supersymmetry algebras (\ref{psusy}).

Solutions of the equations (\ref{q-pereq}) have a quite
complicated structure and general methods of solving such
differential-delay equations give relatively weak results.
E.g., for $N=1$ in \cite{SkS} the existence and uniqueness of solutions
analytical near the $x=0$ point was proved under some constraints,
and in \cite{Liu} existence of the
nonsingular for $x\in\bf R$ solutions was demonstrated (the $|x|\to \infty$
asymptotics of such solutions is not determined yet completely).
Some understanding of the complexity of general solutions is obtained
from considerations of special values of the parameters $q, \mu_j$,
when $f_j(x)$ can be expressed through some known functions.

Let $q$ be arbitrary and $f_j^2(x)=
\frac{1}{1-q^2}\sum_{m=0}^{N-1}\mu_m -\sum_{m=0}^{j-1}\mu_m.$
This yields $L_0=-d^2/dx^2$ or the free quantum mechanical particle,
which acquires in this way a $q$-algebraic interpretation \cite{S2}.

Suppose that $f_0(x)$ is not singular at $x=0$, then in the crystal base
limit $q\to 0$ the potential $u_0(x)$ boils down to the general KdV
$N$-soliton potential.

Substitute now $q=1$ into (\ref{q-pereq}) and assume that
$\sum_{m=0}^{N-1}\mu_m\neq 0$.
Then for $N=1,2$ one easily gets the
potentials $u_0(x)\propto x^2, ax^2+b/x^2$. For $N=3$ the
corresponding system of equations for $f_j(x)$ provides
a ``cyclic" representation of the Painlev\'e-IV equation
\cite{Bu,VS}. The $N=4$ case is related to the Painlev\'e-V
function \cite{Ad}, etc. This gives also a commutator representation
of the Painlev\'e equations which was noticed first in \cite{Fl}.
   For $q=-1$ one gets similar situations
with the functions $f_j(x)$ obeying certain parity symmetry \cite{S2}.
If $\sum_{m=0}^{N-1}\mu_m= 0$ then for odd $N$ and some cases of even $N$
the potential $u_0(x)$ is expressed through the hyperelliptic
functions \cite{VS,Weiss}.

The case when $q$ is a primitive root of unity, $q^n=1,
q\neq \pm1,$ is quite interesting \cite{SkS}. For any odd $n$ and, under
certain conditions, for even $n$ potentials are expressed through
hyperelliptic functions with additional
crystal\-lo\-graphic or quasi-crys\-tal\-lographic symmetries.
  For example, for $N=1$ and $q^3=1$ or $q^4=1$ one recovers the
Lam\'e equation with the equi\-an\-harmonic or lem\-ni\-scatic Weierstrass
$\cal P$-functions. Thus, this classical differential equation of the
second order appears to be related to the representations of the
$q$-harmonic oscillator algebra for $q$ a root of unity.

The general family of self-similar potentials
unifies some of the Painlev\'e functions with hyperelliptic functions.
Due to the connections with quantum algebras, it may be
taken as a new class of nonlinear $q$-special functions (viz. continuous
$q$-Painlev\'e transcendents) defined upon the differential equations.
The standard basic hypergeometric functions \cite{GR}
appear in this formalism through the coherent states.

As far as the discrete spectrum of self-similar potentials is concerned,
from the factorization method and (\ref{q-per}) it follows that
in the simplest case it is composed from $N$ independent geometric
progressions: $\lambda_{pN+k}=\lambda_kq^{2p}, \; k=0,\dots, N-1,\;
p\in {\bf N}$. The same follows from the theory of unitary representations
of the algebra (\ref{q-alg}).
The only condition for the validity of this formal conclusion
is that the functions $f_j(x)$ are non-singular for $x\in \bf R$ and
positive for $x\to +\infty$ and negative for $x\to-\infty$.

\section{Coherent states}

There are many definitions of quantum mechanical coherent states \cite{Per}.
In the context of spectrum generating algebras, such as (\ref{pol}) and
(\ref{q-alg}), they are defined as eigenfunctions of the
corresponding lowering operators. Such eigenfunctions play the
role of generating functions for the space elements of irreducible
representations of a taken algebra. Let us describe briefly
coherent states for the quantum algebra \re{q-alg} (for more details,
see \cite{S2}).

Let us denote as $|\lambda\rangle$ eigenstates of an abstract
operator $L$ entering (\ref{q-alg}),
$L|\lambda\rangle=\lambda|\lambda\rangle.$
Suppose that the operators
$B^\pm$ are conjugated to each other with respect to some
inner product $\langle \sigma|\lambda\rangle$.
We assume that the spectrum of $L$ is not degenerate.
Then the action of operators $B^\pm$ upon $|\lambda\rangle$
has the form:
$$
B^-|\lambda\rangle=
\prod_{k=0}^{N-1}\sqrt{\lambda-\lambda_k}|\lambda q^{-2}\rangle, \qquad
B^+|\lambda\rangle=
\prod_{k=0}^{N-1}\sqrt{\lambda q^2-\lambda_k}|\lambda q^{2}\rangle.
$$
Let $N$ be odd, $0<q^2<1$ and $\lambda_0<\dots <\lambda_{N-1}<q^2\lambda_0<0$.
Then for
$\lambda<0$ the operator $L$ may have a discrete spectrum formed only from
up to $N$ geometric progressions corresponding to the lowest weight unitary
irreducible representations of the algebra \re{q-alg}.
This follows from the fact that $B^-$ is the lowering operator for
the $\lambda<0$ states and $\prod_{k=0}^{N-1}(\lambda-\lambda_k)$
becomes negative for $\lambda< \lambda_0$. Since
$B^-|\lambda_k\rangle=0$, this problem is avoided
for special values of $\lambda$, namely, for
$\lambda=\lambda_{pN+k}\equiv \lambda_kq^{2p}, \; p\in {\bf N}$.
   For normalizable
$|\lambda_k\rangle$, one gets series of normalizable eigenstates
of the form $|\lambda_{pN+k}\rangle \propto
\left(B^+\right)^p|\lambda_k\rangle$.

Coherent states of the first type are defined as eigenstates of $B^-$:
\begin{equation}
B^-|\alpha\rangle_-^{(k)}=\alpha|\alpha\rangle_-^{(k)},
\quad k=0, \dots, N-1,
\label{coh1}\end{equation}
where $\alpha$ is a complex variable.
Representing $|\alpha\rangle_-^{(k)}$ as a sum of the states
$|\lambda_{pN+k}\rangle$ with some coefficients, one finds
$$
|\alpha\rangle_-^{(k)}= \sum_{p=0}^\infty C_p^{(k)}\alpha^p
|\lambda_{pN+k}\rangle =C^{(k)}(\alpha)\; {_N}\varphi_{N-1}\left(
{0, \dots, 0 \atop \lambda_0/\lambda_k, \dots,
\lambda_{N-1}/\lambda_k}; q^2, z\right)|\lambda_k\rangle,
$$
where ${_N}\varphi_{N-1}$ is a standard basic hyper\-geo\-metric series
with the operator argument $z=(-1)^N\alpha B^+/\lambda_0\cdots\lambda_{N-1}$
and $C^{(k)}(\alpha)$ is a normalization constant (in the set of parameters
of the $_N\varphi_{N-1}$ function the term $\lambda_k/\lambda_k=1$
is assumed to be absent).
The superscript $k$ simply counts the number of lowest weight
irreducible representations of the algebra (\ref{q-alg}) each of which
has its own coherent state (or a generating function).
The states $|\alpha\rangle_-^{(k)}$ are normalizable
for $|\alpha|^2< |\lambda_0\cdots\lambda_{N-1}|$. Coherent states of
this type were first constructed for the $q$-harmonic oscillator
algebra (\ref{qosc}) in \cite{AC}, for a particular explicit model
of them, see e.g. \cite{ASu}.

It is not difficult to see, that zero modes of the operator $L$
define a special degenerate representation of the algebra (\ref{q-alg}).
Since $B^\pm$ operators commute with $L$ in the subspace of these
zero modes, one can consider them as coherent states as well.

Unusual coherent states appear from the non-highest
weight representations of (\ref{q-alg}) corresponding to
the $\lambda>0$ eigenstates of $L$.
Let $\lambda_+>0$ be a discrete spectrum
point of $L$. Then the operators $B^\pm$ generate from $|\lambda_+\rangle$
a part of the discrete spectrum of $L$ in the form
of a bilateral geometric progression $\lambda_+ q^{2n}, n\in {\bf Z}$.
Since $\prod_{k=0}^{N-1}(\lambda_+-\lambda_k)>0 $ for arbitrary $N$ and
$\lambda_+>0$, the number of such irreducible representations in
the spectrum of $L$ is not limited. Moreover, a continuous direct sum
of such representations may form a continuous spectrum of $L$.

An important fact is that for $\lambda>0$ the role of lowering operator is
played by $B^+$. Therefore one can define coherent states as eigenstates
of this operator as well \cite{S2}:
\begin{equation}
B^+|\alpha\rangle_+=\alpha|\alpha\rangle_+.
\label{coh2}\end{equation}
Suppose that the bilateral sequence $\lambda_+q^{2n}>0$ belongs to the
discrete spectrum of $L$. Then the states $|\alpha\rangle_+$ are expanded
in the series of eigenstates $|\lambda_+q^{2n}\rangle$:
$$
|\alpha\rangle_+= \sum_{n=-\infty}^\infty C_n\alpha^n|\lambda_+ q^{2n}
\rangle =C(\alpha)\; {_0\psi_N}\left({0, \dots, 0\atop \lambda_0/\lambda_+, \dots,
\lambda_{N-1}/\lambda_+}; q^2, z \right)|\lambda_+\rangle,
$$
where ${_0\psi_N}$ is a bilateral $q$-hypergeometric series
with the operator argument $z=\alpha B^-/(-\lambda_+)^N$
and $C(\alpha)$ is a normalization constant.
In this case the states $|\alpha\rangle_+$ are normalizable if
$|\alpha|^2>|\lambda_0\cdots\lambda_{N-1}|$.

Suppose that the $\lambda>0$ region is occupied by the continuous spectrum.
The corresponding states $|\lambda\rangle$ may be normalized
as $\langle \sigma|\lambda\rangle=\lambda\delta(\lambda-\sigma)$.
Let $N=1$ and the states $|\lambda\rangle$ are not degenerate.
Taking $\rho=1$ in (\ref{qosc}), we get the expansion
$$
|\alpha\rangle_+^{(s)}= C(\alpha)
\int_0^\infty \frac{\lambda^{\gamma_s}|\lambda\rangle d\lambda}
{\sqrt{(-\lambda q^2(1-q^2);q^2)_\infty}},
$$
where
$$
\gamma_s=\frac{2\pi i s-\ln(\alpha q^2\sqrt{1-q^2})}
{\ln q^2}, \qquad   s\in {\bf Z}.
$$
There is a countable family of such coherent states which
have a unit norm for $|\alpha|^2 > 1/(1-q^2)$ under the
following choice of the normalization constant $C(\alpha)$:
$$
|C(\alpha)|^{-2}=\int_0^\infty\frac{\lambda^{-\nu}d\lambda}
{(-\lambda q^2(1-q^2);q^2)_\infty}
=\frac{\pi}{\sin\pi\nu}\frac{(q^{2\nu};q^2)_\infty
(q^2(1-q^2))^{\nu-1}}{(q^2;q^2)_\infty},
$$
where $\nu=\ln|\alpha q\sqrt{1-q^2}|/\ln q.$
The integral under consideration is a particular subcase of
a Ramanujan $q$-beta integral \cite{GR}.

We were considering until now the abstract coherent states.
In the case of Schr\"odinger equations with the self-similar potentials
one has the doubly degenerate
continuous spectrum for $\lambda>0$. The structure of $|\alpha\rangle_\pm$
states in this case is very complicated. Even for the zero potential
case (the free particle) we get a highly nontrivial situation.
Let $L=-d^2/dx^2$ and
$$
B^-=U^{-1}(d/dx+1/\sqrt{1-q^2}),\qquad B^+=(-d/dx+1/\sqrt{1-q^2})U,
$$
where $U$ is the dilation operator, $Uf(x)=q^{1/2}f(qx), \; 0<q<1.$
It can be checked directly that $B^-B^+-q^2B^+B^-=1$ and
$L=B^+B^- - 1/(1-q^2).$ The equation $B^-\psi_\alpha^-(x)=
\alpha\psi_\alpha^-(x)$ coincides now with the retarded pantograph equation,
which was investigated in detail in \cite{KaM}.
As follows from the results of this paper, in the retarded case the
pantograph equation does not admit solutions belonging to $L^2({\bf R})$.
However, the operator $B^+$ has infinitely many normalizable
eigenstates. The equation $B^+\psi_\alpha^+(x)=\alpha\psi_\alpha^+(x)$
coincides now with the advanced pantograph equation
\begin{equation}\label{pant}
\frac{d}{dx}\psi_\alpha^+(x)=-\alpha q^{-3/2}\psi_\alpha^+(q^{-1}x)
+\frac{q^{-1}}{\sqrt{1-q^2}}\psi_\alpha^+(x),
\end{equation}
which has infinitely many solutions belonging to $L^2({\bf R})$ \cite{KaM}.
   For the free particle realization of the $N>1$ symmetry algebras,
one gets the generalized pantograph equations considered in \cite{Is}.
As one of the open problems in this field, let us mention a need
to characterize a minimal set of solutions of the equation (\ref{pant})
providing a complete basis of the Hilbert space $L^2({\bf R})$.

\section{Solitons, Ising chains, and random matrices}

Consider the KdV $N$-soliton solution $u_N(x, t)=- 2\partial_x^2
\log \tau_N(x, t),$  where $\tau_N$ is a Wronskian of $N$
different solutions of the free Schr\"odinger equation (\ref{crum}).
Actually, there are many determinantal representation for this
tau-function, e.g. \cite{AS}
\be
\tau_N=\det \Phi, \qquad \Phi_{ij}=\delta_{ij}+ {2\sqrt{k_i k_j}
\over k_i+k_j} e^{(\theta_i+\theta_j)/2},
\lab{phase}\ee
$$
\theta_i=k_ix-k_i^3 t +\theta_i^{(0)}, \qquad i, j=1, 2, \dots, N.
$$
The variables $k_i$ are known to describe amplitudes of solitons
(they are related to the eigenvalues of $L_N$ as $\lambda_i=-k_i^2/4$),
$\theta_i^{(0)}/k_i$ are the $t=0$ phases of solitons
and $k_i^2$ are their velocities.

A subclass of the self-similar potentials appearing from the $q$-periodic
closure (\ref{q-per}) can be viewed as a particular infinite soliton
potential. Indeed, consider the KdV $pM$-soliton solution with the
parameters $k_j$ subject to the constraint $k_{j+M}=qk_j, 0<q<1$, and
take the $p\to\infty$ limit. The limiting potential has
the self-similar spectrum $k_{pM+m}=q^pk_m, m=0, \dots, M-1,$
and an infinite number of free parameters $\theta_j^{(0)}$.

The scaled potential $\tilde u(x,t)=q^2u(qx,q^3t)$
has the same solitonic interpretation with the phases
$\tilde \theta_j(x,t)=$ $\theta_j(qx,q^3t)=$
$k_{j+M}x-k_{j+M}^3t+\theta_j^{(0)}.$
Let us demand that $\theta_{j+M}^{(0)}=\theta_j^{(0)}$, i.e.
impose the additional constraint $\theta_j(qx, q^3 t)=
\theta_{j+M}(x,t)$, which is seen to match with the condition (\ref{q-per}).
In this picture the discrete dilation $x\to qx, t\to q^3t$ just deletes
$M$ solitons corresponding to the smallest eigenvalues of $L$ --- this
feature is the simplest physical characterization of the
self-similar potentials.

Let us turn now to the statistical mechanics applications.
$N$-soliton tau-function of the KdV equation \re{phase} can be represented
in the following Hirota form \cite{AS,Hir}:
\begin{equation}
\tau_N = \sum_{\sigma_i=0,1} \exp \left( \sum_{0\leq i<j \le N-1 } A_{ij}
\sigma_i \sigma_j + \sum_{i=0}^{N-1} \theta_i\sigma_i \right).
\label{N_soliton}\end{equation}
The coefficients $A_{ij}$, describing phase shifts of the $i$-th
and $j$-th solitons after their collision, have the form
\begin{equation}
e^{A_{ij}}={ (k_i-k_j)^2 \over  (k_i+k_j)^2 }.
\label{KDV_phase}\end{equation}
As noticed in \cite{LS}, for $\theta_i=\theta^{(0)}$
this tau-function coincides with the grand
partition function of a lattice gas model on a line
(for a two-dimensional Coulomb gas picture, see \cite{LS2}
and the next section). In this interpretation the discrete variables
$\sigma_i$ describe filling factors of the lattice sites by molecules,
$\theta^{(0)}$ is a chemical potential, and $A_{ij}$ are the interaction
energies of molecules \cite{Bax}.

It is well known that a simple change of variables in the lattice
gas partition function leads to the partition function
of some Ising chains:
\begin{equation}\label{Zising}
Z_N=\sum_{s_i=\pm1} e^{- \beta E}, \qquad
E =\sum_{0\leq i<j \leq N-1} J_{ij}s_is_j -\sum_{i=0}^{N-1} H_i s_i,
\end{equation}
where $N$ is the number of spins $s_i=\pm 1$, $J_{ij}$ are  the
exchange constants, $H_i$ is an external magnetic field, $\beta=1/kT$
is the inverse temperature. Indeed, a replacement of filling factors
in \re{N_soliton} by the spin variables $\sigma_i=(s_i+1)/2$ yields
\be
\tau_N=e^{\varphi} Z_N, \qquad \varphi=\frac{1}{4}\sum_{i<j} A_{ij}
+ \frac{1}{2}\sum_{j=0}^{N-1} \theta_j,
\lab{tauising}\ee
where
\be
\beta J_{ij} =-\frac{1}{4} \; A_{ij},
\qquad
\beta H_i=\frac{1}{2}\theta_i +\frac{1}{4}\sum_{j=0, i\neq j}^{N-1} A_{ij}.
\lab{identif}\ee

Similar relations with Ising chains are valid for the whole infinite
Kadomtsev-Petviashvili (KP) hierarchy of equations
and some other differential and finite difference nonlinear integrable
evolution equations. Corresponding tau-functions have the same form
\re{N_soliton} with a more complicated structure of the phase
shifts $A_{ij}$ and phases $\theta_i$ (a partial list of such
equations can be found in \cite{AS,Hir}).

The general relation between $N$-soliton solutions of integrable hierarchies
and partition functions of the lattice statistical mechanics models
established in \cite{LS} brings a new point of view upon the self-similar
potentials as well. Namely, as shown in \cite{LS} self-similar
spectra can be deduced from the condition of translational invariance of
the infinite spin chains. E.g., let us demand that the spin chain is
invariant with respect to the shift of the lattice by one site
$j\to j+1$ which assumes that $J_{i+1,j+1}=J_{ij}$.
As a result, the exchange constants $J_{ij}$ (or the soliton phase
shifts $A_{ij}$) depend only on the distance between the sites $|i-j|$.
This very simple and quite natural physical requirement forces $k_i$
to form one geometric progression (N.B.: uniquely)
\begin{equation}
k_i=k_0q^i, \qquad q=e^{-2\alpha}, \qquad A_{ij}=2\ln|\tanh\alpha (i-j)|,
\label{interactions} \end{equation}
where $k_0$ and $0<q<1$ are some free parameters.
In a more complicated case one demands the translational invariance
with respect to the shifts by $M$ lattice sites, i.e.
$J_{i+M,j+M}=J_{ij}$, and this results in the general
self-similar spectra $k_{j+M}=qk_j$.
The main drawback of these relations between the solitons and Ising chains
is that for fixed $q$ the temperature $T$ (or $\beta$) is fixed too,
which is clearly seen from the comparison of (\ref{identif}) and
(\ref{interactions}) (in the Coulomb gas interpretation one actually finds
that $\beta=2$).

  For finite chains $0\leq j\leq N-1$, the translational invariance is
not exact.
The infinite soliton limit $N\to\infty$ corresponds to the thermodynamical
limit in statistical mechanics. In this picture, all coordinates of
integrable hierarchies ($x,t$ and higher hierarchy ``times") are interpreted
as particular parameters of the external magnetic field $H_i$.
Since $0<q<1$, the $x$ and $t$-dependent part of $H_i$ decays exponentially
fast for $i\to\infty$. As a result, in the limit $N\to\infty$
only the constants $\theta_i^{(0)}$ are relevant for the
partition function (speaking more precisely, they determine the
leading asymptotics term of the partition function at $N\to\infty$).
Although in the thermodynamic limit the $x,t$-dependence is
washed out and the Ising chain is periodic, presence of an infinite
number of free parameters $\theta_i^{(0)}$ does not allow one to find
a closed form
expression for the leading term of $Z_N$. However, for the self-similar
infinite soliton systems, characterized by the periodicity
of $\theta_i^{(0)}$ or, equivalently, by the periodicity of the
external magnetic field $H_{i+M}=H_i$, such expressions do exist.

Take $M=1$ and let the magnetic field be homogeneous $H_i=H$.
Then in the KdV equation case one gets an anti\-ferro\-magnetic
Ising chain: $0<|\tanh \alpha(i-j)|<1$ and $J_{ij} = -A_{ij}/4\beta > 0$
(a similar picture is valid for the general $M$).
The exchange has a long distance character but its intensity falls
off exponentially fast and, as a result, phase transitions
are absent for nonzero temperatures. However, for $\alpha\to 0$ or $q\to 1$
the exchange constants $J_{ij}$ are diverging. Under appropriate
renormalizations one can build a model with a very small non-local
exchange such that a phase transition takes place even for
non-zero $T$.

In order to compute the partition function one can use determinantal
representations for the tau-function, e.g. the Wronskian formula
(\ref{crum}). As shown in \cite{LS}, for the translationally invariant
Ising chains in a homogeneous magnetic field tau-functions become
determinants of some Toeplitz matrices and their natural $M$-periodic
generalizations. E.g., for $M=1$ the following result has been derived:
$Z_N\to \exp(-N\beta f_I)$ for $N\to\infty$,
where the free energy per site $f_I$ has the form
\be
-\beta f_I(q, H)=\ln \frac{2(q^4;q^4)_\infty \cosh \beta H}
{(q^2;q^2)_\infty^{1/2}}
+ \frac{1}{4\pi}\int_0^{2\pi}d\nu \ln (|\rho(\nu)|^2 - q\tanh^2 \beta H),
\lab{free}\ee
$$
|\rho(\nu)|^2=
\frac{(q^2e^{i\nu};q^4)_\infty^2(q^2e^{-i\nu};q^4)_\infty^2}
{(q^4e^{i\nu};q^4)_\infty^2(q^4e^{-i\nu};q^4)_\infty^2}\;
\frac{1}{4\sin^2(\nu/2)}=
q\frac{\theta_4^2(\nu/2, q^2)}{\theta_1^2(\nu/2, q^2)}
$$
($\theta_{1,4}$ are the standard Jacobi theta-functions).
In the derivation of this formula the Ramanujan $_1\psi_1$ sum
was used at the boundary of convergence of the corresponding
infinite bilateral basic hypergeometric series. A standard
physicists' trick was used for dealing with that, namely, a
small auxiliary parameter $\epsilon$ was introduced into
the original expression for the density function $\rho(\nu)$
\begin{equation}
\rho(\nu)=\frac{(q^2;q^4)_\infty^2}{(q^4;q^4)_\infty^2}
\sum_{k=-\infty}^\infty
\frac{e^{i\nu k-\epsilon k}}{1-q^{4k+2}}=
\frac{(q^2e^{i\nu-\epsilon};q^4)_\infty (q^2e^{-i\nu+\epsilon};q^4)_\infty}
{(e^{i\nu-\epsilon};q^4)_\infty (q^4e^{-i\nu+\epsilon};q^4)_\infty},
\label{nu}
\end{equation}
which guarantees the absolute convergence of the series for $\epsilon> 0$.
Although the limit $\epsilon\to 0$ leads to some singularity in $\nu$
on the right hand side of (\ref{nu}), it does not lead to divergences
after taking the integral in (\ref{free}) (a more precise
mathematical justification of this physical ``harm\-less\-ness"
is desirable). The total magnetization of the lattice
$m(H)=-\partial_H f_I = \stackreb{\lim}{N\to\infty} N^{-1}
\sum_{i=0}^{N-1}\langle s_i \rangle$ has the following appealing form
\be
m(H)=\left(1-\frac{1}{\pi}
\int_0^{\pi}\frac{\theta_1^2(\nu, q^2)d\nu}{\theta_4^2(\nu, q^2)\cosh^2\beta H
-\theta_1^2(\nu, q^2) \sinh^2\beta H} \right)\tanh \beta H.
\lab{magkdv}\ee

In order to change the fixed value of the temperature, one may
try to replace (\ref{interactions}) by $A_{ij}=2 n\ln|(k_i-k_j)/(k_i+k_j)|,$
where $n$ is some sequence of integers, and look for integrable equations
admitting $N$-soliton solutions with such phase shifts.
This is a highly nontrivial task, and in \cite{LS}
only one more permitted value of the temperature was found.
It corresponds to $n=2$ and appears from special reduction of the
$N$-soliton solution of the KP-equation of B-type (BKP) \cite{DJKM}.
The general BKP $\tau$-function generates much more complicated Ising
chains than in the KdV case. The corresponding exchange
constants have the form
\begin{equation}
\beta J_{ij}=-\frac{1}{4}A_{ij}, \quad e^{A_{ij}}=\frac{(a_i-a_j)(b_i-b_j)(a_i-b_j)(b_i-a_j)}
{(a_i+a_j)(b_i+b_j)(a_i+b_j)(b_i+a_j)},
\label{BKP}\end{equation}
where $a_i,b_i$ are some free parameters.
Translational invariance of the spin lattice, $J_{ij}=J(i-j)$,
results in the following spectral self-similarity
\begin{equation}
a_i=q^i, \quad b_i=bq^i, \quad q=e^{-2\alpha},
\label{BKP_momentums}
\end{equation}
where we set $a_0=1$ and assume that $\alpha >0$.
This gives the exchange
$$
\beta J_{ij}=-\frac{1}{4}\ln\frac{\tanh^2\alpha(i-j)- (b-1)^2/(b+1)^2}
{\coth^2\alpha(i-j) - (b-1)^2/(b+1)^2}.
$$
The parameter $b$ is restricted to the unit disk $|b|\leq 1$ due to
the inversion symmetry $b\to b^{-1}$. For $-1 < b < -q$ one gets now
the ferromagnetic Ising chain, i.e. $J_{ij}<0$;
for $q< b \leq 1$ and $b=e^{i\phi}\neq -1$
the chain is antiferromagnetic, i.e. $J_{ij}>0$.
In the thermodynamic limit $N \to \infty $ the partition function
$Z_N$ is given again by the determinant of a Toeplitz matrix,
which is diagonalized by the discrete Fourier transformation.
Similar picture holds for arbitrary $M$-periodic chains,
see \cite{LS} for explicit expressions for the magnetization and
technical details of computations.

It is well known that the $n$-particle lattice gas partition functions
define probability distribution functions in the theory
of random matrices. In a natural way, partition functions
of the Ising models considered in the previous section may be
related to the probability distribution functions of some random
matrix models with a discretized set of eigenvalues. Let us
describe briefly this correspondence comparing the KP $N$-soliton solution
and the Dyson's circular ensemble.

Dyson has introduced an ensemble of unitary random $n\times n$
matrices with the eigenvalues $\epsilon_j=e^{i\phi_j}$, $j=1, \dots, n$,
such that after the integration over the auxiliary ``angular variables"
the probability distribution of the phases $0\leq \phi_j<2\pi$
takes the form $Pd\phi_1 \dots d\phi_n\propto
\prod_{i<j}|\epsilon_i-\epsilon_j|^2 d\phi_1 \dots d\phi_n.$
One may relax some of the conditions used by Dyson and work with a more
general set of distribution functions. In \cite{Gaud1} Gaudin has proposed
a circular ensemble with the probability distribution law
\begin{equation}
Pd\phi_1 \dots d\phi_n\propto
\prod_{i<j}\left|\frac{\epsilon_i-\epsilon_j}
{\epsilon_i-\omega\epsilon_j}\right|^2 d\phi_1 \dots d\phi_n,
\label{distr1}
\end{equation}
containing a free continuous parameter $\omega$. It
interpolates between the Dyson's case  ($\omega=0$) and the uniform
distribution ($\omega=1$). Let us note in passing, that the paper
\cite{Gaud1} contains in it implicitly a specific model of the
$q$-harmonic oscillator. The model (\ref{distr1}) admits also
an interpretation as a lattice gas on the circle with the partition function
$$
Z_n\propto \int_0^{2\pi}\dots\int_0^{2\pi}d\phi_1\dots d\phi_n
\exp\left(-\beta \sum_{i<j}V(\phi_i-\phi_j)\right),
$$
where $\beta=1/kT=2 $ is fixed and the potential energy has the form
\begin{equation}
\beta
V(\phi_i-\phi_j)=
\ln\left(1+\frac{\sinh^2\gamma}{\sin^2((\phi_i-\phi_j)/2)}\right),
\qquad \omega=e^{-2\gamma}.
\label{gas1}
\end{equation}

A surprising fact is that the grand partition function of this model
may be obtained in a special infinite soliton limit of the $N$-soliton
tau-function of the KP hierarchy \cite{LS}. This means that the finite
KP soliton solutions provide a discretization of the model, namely,
they define a lattice gas on the circle. This leads to random matrices
with a discrete set of eigenvalues. E.g., one may take unitary
$n\times n$ matrices with eigenvalues equal to $N$-th roots of unity, i.e.
$\epsilon_j=\exp(2\pi im_j/N),\; m_j=0, \dots, N-1 $.
The probability measure is taken to be continuous in the auxiliary
``angular" variables of the unitary matrices and discrete in the
eigenvalue phase variables $\phi_j$.
More precisely, the integrals over $\phi_j$ are replaced by finite sums
over $m_j$ and the continuous model is recovered for $m_j, N\to\infty$
with finite $m_j/N$:
\begin{equation}
\left(\frac{2\pi}{N}\right)^n\sum_{m_1=0}^{N-1} \dots \sum_{m_n=0}^{N-1}
\stackreb{\to}{N\to\infty} \int_{0}^{2\pi}d\phi_1 \dots
\int_{0}^{2\pi}d\phi_n.
\label{sum}\end{equation}
The $n$-particle partition function becomes
$$
Z_n(N,\omega)=\left(\frac{2\pi}{N}\right)^n\sum_{m_1=0}^{N-1} \dots
\sum_{m_n=0}^{N-1} \prod_{1 \le i < j \le n}
\left|\frac{\epsilon_i-\epsilon_j}
{\epsilon_i/\sqrt{\omega}-\sqrt{\omega}\epsilon_j}\right|^2,
$$
while the grand canonical ensemble partition function takes the form
\begin{equation}
Z(\omega,\theta)=\sum_{n=0}^N \frac{Z_n(N,\omega)e^{\theta n}}{n!} =
\sum_{\sigma_m=0,1} \exp \left( \sum_{0 \le m<k \le N-1}A_{mk}
\sigma_m\sigma_k+(\theta+\eta) \sum_{m=0}^{N-1} \sigma_m \right),
\label{grand}\end{equation}
where $\eta=\ln({2\pi}/{N})$ is an excessive chemical potential, and
$$
A_{mk} = \ln \frac{\sin^2(\pi(m-k)/N)}{\sin^2(\pi(m-k)/N) +\sinh^2\gamma}
= \ln \frac{(a_m-a_k)(b_m-b_k)}{(a_m+b_k)(b_m+a_k)},
$$
are the KP solitons phase shifts for a special choice of the parameters
\begin{equation}
a_m=e^{2\pi i m/N}, \qquad b_m=-\omega a_m, \quad m=0,1, \dots, N-1.
\label{momentums}\end{equation}

To conclude, the grand partition function of the Gaudin's
circular ensemble coincides with the particular infinite
soliton KP tau-function at zero hierarchy ``times".
The root of unity discretization of circular ensembles has been
considered by Gaudin himself \cite{Gaud2}, where a connection with
Ising chains was noticed as well, but the relation with soliton
solutions of integrable equations was not established.
The BKP hierarchy of equations suggests a generalization of
the distribution law (\ref{distr1}) to
\begin{equation}
Pd\phi_1 \dots d\phi_n\propto
\prod_{i<j}\left|\frac{\epsilon_i-\epsilon_j}
{\epsilon_i+\epsilon_j}\right|^2\left|\frac{\epsilon_i+\omega\epsilon_j}
{\epsilon_i-\omega\epsilon_j}\right|^2 d\phi_1 \dots d\phi_n,
\label{distribution2}\end{equation}
and its discrete analogue, which were not investigated yet appropriately.

It is well known that the $n$-tuple Selberg integral provides an explicit
evaluation of the $n$ particle Coulomb gas partition functions for arbitrary
values of the inverse temperature $\beta$ (i.e. not just for $\beta=1,2,4,$
corresponding to orthogonal, unitary and symplectic ensembles).
It would be interesting to know whether a similar
universal formula exists for the grand partition functions of
lattice Coulomb gases.

\section{Lattice Coulomb gas on the plane}

Solution of the Poisson equation on the plane,
$\Delta V({\bf r},{\bf r}^\prime)= - 2\pi\delta({\bf r}-{\bf r}^\prime),$
defines the electrostatic potential $V(z,z^\prime)=-\ln |z-z^\prime|$,
created by a charged particle placed at the point $z'=x'+iy'$.
In the bounded domains with dielectric or conducting walls, the
potential has a more complicated form since the normal component of the
electric field ${\cal \bf E} = -{\bf \nabla} V$ should vanish on the surface
of a dielectric, ${\cal E}_n = 0,$ while the tangent component is zero
at the metallic boundary, ${\cal E}_t =0.$ For simple geometric
configurations of boundaries an introduction of artificial images of
charges may simplify solution of the Poisson equation.

The energy of an electrostatic system (``plasma") consisting of $N$
particles in a bounded domain of the plane is
\begin{equation}
E_N=\sum_{1\le i<j\le N} q_iq_jV(z_i,z_j)+\sum_{1\le i\le N} q_i^2v(z_i)
+\sum_{1\le i\le N} q_i\phi(z_i),
\label{energy}
\end{equation}
where $z_j=x_j+i y_j$ and $q_j$ are the particles' coordinates and charges.
The first term is the standard Coulomb energy, the second one describes an
interaction with the boundaries (or the charge-image interaction),
and the last term corresponds to the external fields contribution.

Suppose that our plasma is composed from the particles of equal charges,
$q_j=+1$ (in the two component case $q_j=\pm 1$), upon a discrete lattice
$\Gamma$ on the plane. The grand partition function of such lattice
Coulomb gas is
$$
G_N=\sum_{n=0}^{N}\frac{e^{\mu n}}{n!}\sum_{z_1\in \Gamma}
\dots \sum_{z_n\in \Gamma}  e^{-\beta E_n},
$$
where $\mu$ is an effective chemical potential.
In general case one can rewrite $G_N$ in the form \cite{Bax}
\begin{equation}\label{G}
G_N=\sum_{\{\sigma(z)\}}\exp\Bigl(\frac{1}{2}
\sum_{z\neq z^\prime}W(z,z^\prime)\sigma(z)\sigma(z^\prime)
+\sum_{z\in \Gamma}w(z)\sigma(z)\Bigr),
\end{equation}
$$
W\left(z,z^\prime\right)=-\beta q(z)q(z^\prime)V(z,z^\prime), \quad
w(z)=\mu(z)-\beta\left(q^2(z)v(z)+q(z)\phi(z)\right),
$$
where $\sigma(z)=0$ or 1 depending on whether the site with the
coordinate $z$ is empty or occupied by a particle,
and the functions $q(z)$ and $\mu(z)$ characterize distribution of
particles of different types. E.g., in the two-component case,
when $q({z_\pm})={\pm 1}$ charges occupy the $\{z_\pm\}$ sublattices,
one has $\mu({z_\pm})=\mu_{\pm}$. Note that these
2D lattice gases describe simultaneously some 2D Ising magnets.

Let us take now the $N$-soliton tau-function (\ref{N_soliton})
and replace in it the soliton number $j$ by a variable $z$ taking
$N$ discrete values. As a result, it may be rewritten in the form
\begin{equation}
\tau_N=\sum_{\sigma(z)=0,1}\exp\Bigl(\frac{1}{2}\sum_{z\neq z^\prime}A_{z
z^\prime}\sigma(z)\sigma(z^\prime)+\sum_{\{z\}}
\theta(z)\sigma(z)\Bigr).
\label{tau}\end{equation}
But this is precisely (\ref{G}) --- one just needs to identify the phase
shifts $A_{zz^\prime}$ with the Coulomb interaction potential
$W(z,z^\prime)$, and the phases $\theta(z)$ with the function $w(z)$.
As a result of this observation made in \cite{LS2}, one can construct
a number of new solvable models of Coulomb gases in addition to already
known ones (see, e.g. \cite{FJT,Gaud1,Gaud2} and references therein).

    For the KP-hierarchy soliton solutions one has \cite{DJKM}
\begin{equation}
A_{zz^\prime}=\ln\frac{(a_z-a_{z^\prime})(b_z-b_{z^\prime})}
{(a_z+b_{z^\prime})(b_z+a_{z^\prime})}, \qquad
\theta(z)=\theta^{(0)}(z)+\sum_{p=1}^\infty(a_z^p-(-b_z)^p)t_p,
\label{kp}\end{equation}
where $t_p$ is the $p$-th KP hierarchy ``time"
and $a_z, b_z$ are some arbitrary functions of $z$.
    For the BKP-hierarchy $A_{zz'}$ are given by (\ref{BKP}) and
$\theta(z)=\theta^{(0)}(z)+\sum_{p=1}^\infty(a_z^{2p-1}+b_z^{2p-1})t_{2p-1},$
where $t_{2p-1}$ are the BKP evolution ``times".

Let us take $a_z=z=x+i y, \; b_z=-z^*=-x +i y.$ Then in the KP case
\begin{equation}\label{kpint}
A_{zz'}=W(z,z^\prime)=-2 V(z,z')=
2\ln |z-z^\prime| -2\ln |z^*-z^\prime|,
\end{equation}
where $V(z,z')$ is the potential created by a positive unit charge placed
at $z'$ over a conductor with its surface occupying the $y= 0$ line.
In this case $V(z,z')$ solves the Poisson equation with
the boundary condition ${\cal E}_t(y=0)=0$.
The same potential is created by a positive charge placed at the point
$z'$ and its image of opposite charge located at the point
$\left(z'\right)^*$.

Similar to the random matrix models situation, the correspondence
between 2D lattice Coulomb gases with specific boundary conditions
and solitons is valid only for fixed $\beta$, which is found
from the comparison of (\ref{kpint}) with (\ref{G}) to be $\beta=2.$
Since $w(z)=\theta(z)$, one finds an explicit expression for the
zero time soliton phases $\theta^{(0)}(z)$ in (\ref{kp}):
\begin{equation}
\theta^{(0)}(z)=\mu-\beta(\ln |z^*-z|+ \phi(z)),\qquad \beta=2,
\label{fkp}
\end{equation}
where the middle term corresponds to the charge-image interaction, and
$\phi(z)$ is the potential created by a neutralizing background of
some density $\rho({\bf r})$, $\Delta\phi({\bf r})=-2\pi\rho({\bf r}),
\; \phi_x(y=0)=0.$
The harmonic term $\sum_{p=0}^\infty (z^p-(z^*)^p)t_p=-\beta \phi_{ext}(z)$
in (\ref{kp}) corresponds to an external electric field.
One may conclude that the imaginary hierarchy times evolution of the KP
soliton solutions describes the behavior of a 2D lattice Coulomb gas in a
varying external electric field.

Usefulness of the conformal transformations $z\to f(z)$ is well known
in the potential theory. With their help one can map plasma particles to
various regions. So, the map $z\to z^n$ puts them inside a corner with
the $\pi/n$ angle between the conducting walls.
The exponential map $z\to e^{\pi z/L}$ leads to a gas inside the strip
$0< \Im z < L$ between two parallel conductors, etc.

As to the BKP equation case, the choice $a_z=z, b_z=z^*$ leads to the
Coulomb gas placed inside the upper right quarter of the
plane with the dielectric and conductor walls along the $x$ and
$y$-axes respectively. The discrete temperature appears
to be the same, $\beta=2$. A curious set of solvable dipole gas models
is generated by the self-similarity constraints imposed upon the
spectral data of the corresponding multi-soliton systems.
   For further details, see \cite{LS2}.

\section{Appendix. A heuristic guide to special functions
of one variable}

There are many handbooks and textbooks  on special functions.
However, none of them contains a formal list of properties
which a function should have in order to be ``special".
One usually talks on functions of some particular type
(hypergeometric, $L$-functions, etc). R. Askey has suggested
one universal definition that if a function is so useful that
it starts to bear some name, then it is ``special".
Another essential, but not so universal, characteristic
refers to the asymptotic behavior. Namely, for special
functions one is expected to be able to deduce asymptotics at infinity
from a known local behavior, i.e. the connection
problems should be solvable --- such an approach is advocated by people
working on the Painlev\'e type functions.

To the author's taste these two definitions are relying upon
the secondary features of special functions. One has to have already
the functions in hands in order to start to investigate their properties.
If one takes as a goal the search of new special functions, then
it is necessary to find a definition containing a more
extended set of technical tools for work. In this respect it
should be stressed that even the term ``classical special functions"
appears to be not so stable. E.g., it is by now accepted that
the family of classical orthogonal polynomials consists of not just
the Jacobi polynomials and their descendants, but of the essentially
more complicated Askey-Wilson polynomials invented just
two decades ago.

The group theory provides a number of tools for building new
functions, but, unfortunately, connections with the
representation theory often provide only interpretations for
functions already defined by other means. Still, the symmetry approach
seems to be the central one in the theory of special functions.
In particular, the key ``old" special functions appear from
separations of variables in the very simple (and, so,
useful and universal) equations \cite{Mi2}.
   For the last decade the author's research was tied to the
following working definition: {\it special functions are the functions
associated with self-similar reductions of spectral transformation
chains for linear spectral problems.}
Speaking differently, special functions are connected
with fixed points of various continuous and discrete symmetry
transformations for a taken class of eigenvalue problem.
This definition works well only for special functions of one
independent variable (which may, however, depend on an infinite
number of parameters) and even for them it does not pretend
to cover all possible special functions.
On the one hand, this definition comes from the theory of completely
integrable systems, where a search of self-similar solutions of nonlinear
evolution equations is a standard problem \cite{AS}.
On the other hand, particular examples found from this approach
show that it has in its heart the {\it contiguous relations} ---
linear or nonlinear equations
connecting special functions at different values of their parameters.

Schematically, a heuristic algorithm of looking for these ``spectral"
special functions consists of the following steps:

1. Take a linear eigenvalue problem determined by differential or
difference equations.

2. Consider another linear equation involving variables
   entering the first equation in a different way and
   having the same space of solutions.

3. Resolve compatibility conditions of the taken linear problems
and derive nonlinear equations for free functional coefficients
entering these problems. When the second equation is
differential, one gets the continuous flows associated with the
KdV, KP, Toda, etc equations. When the second equation is discrete,
one gets a sequence of Darboux transformations performing some
discrete changes in the spectral data and providing some
discrete-time Toda, Volterra, etc chains.

4. Analyze discrete and continuous symmetries of the latter equations
in the Lie group-theoretical sense, i.e. look for nontrivial
continuous and discrete transformations mapping solutions
of these nonlinear equations into other solutions.

5. Consider {\it self-similar} solutions of the derived nonlinear equations,
which are invariant with respect to taken symmetry transformations.
As a result, some finite sets of nonlinear differential,
differential-difference, difference-difference, etc equations
are emerging which define ``nonlinear" special functions.
Solutions of the original linear equations with
coefficients determined by these self-similar functions
define ``linear" special functions.

The last two steps are heuristic since, despite of a big
progress in the general theory of symmetry reductions
(see, e.g. \cite{LW,Mae}), no completely regular method has been built yet.
E.g., the reductions used in the derivation of the associated Askey-Wilson
polynomials' recurrence coefficients \cite{SZ}
and in the discovery of new explicit systems of biorthogonal rational
functions described in Zhedanov's lecture at this meeting
did not find yet a purely group-theoretical setting.

Another important ingredient of the theory of special functions, which
was not listed in this scheme, is the theory of transcendency. Painlev\'e
functions are known to be transcendental over the
differential fields built from a finite tower of Picard-Vessiot
extensions of the field of rational functions. In solving differential
(difference, or whatever) equations one has to determine
eventually which differential field a taken solution belongs to
(e.g., whether it belongs to the field of functions over which the
differential equation is defined). As an open problem in this field,
which was discussed partially in \cite{S2}, we mention a need to find a
differential (or difference) Galois theory interpretation of the
self-similar reductions of factorization chains.

\medskip
{\bf Acknowledgements.}
The author was partially supported by the RFBR grants No. 00-01-00299 and
00-01-10564.

\end{document}